\documentclass{PoS}

\title{Origin and Evolution of the Multi--band Variability in the Flat Spectrum Radio Source 4C~38.41}

\ShortTitle{Origin and Evolution of Multi--band Variability in 4C~38.41}

\author {\speaker{Juan~Carlos Algaba}$^{a,b}$, {Sang~Sung Lee}$^{b,c}$, {Bindu Rani}$^{d}$, {Dae-Won Kim}$^{a}$, {Motoki Kino}$^{e,f}$, 
{Jeffrey Hodgson}$^{b}$, {Guang-Yao Zhao}$^{b}$, {Do-Young Byun}$^{b}$, {Mark Gurwell}$^{g}$, {Sin-Cheol Kang}$^{b,c}$, {Jae-Young Kim}$^{h}$, {Jeong-Sook Kim}$^{e}$, {Soon-Wook Kim}$^{b,c}$, {Jongh-Ho Park}$^{a}$, {Sascha Trippe}$^{a}$
and Kiyoaki Wajima$^{b}$\\
\llap{$^a$}Department of Physics and Astronomy, Seoul National University, 1 Gwanak-ro, Gwanak-gu, Seoul 08826, Republic of Korea\\
\llap{$^b$}Korea Astronomy \& Space Science Institute, 776, Daedeokdae-ro, Yuseong-gu, Daejeon, 305-348, Republic of Korea\\
\llap{$^c$}Korea University of Science and Technology, 217 Gajeong-ro, Yuseong-gu, Daejeon 34113, Republic of Korea\\
\llap{$^d$}NASA Goddard Space Flight Center, Greenbelt, MD 20771, USA\\
\llap{$^e$}National Astronomical Observatory of Japan, 2211 Osawa, Mitaka, Tokyo 1818588, Japan\\
\llap{$^f$}Kogakuin University, Academic Support Center, 2665-1 Nakano, Hachioji, Tokyo 192-0015, Japan\\
\llap{$^g$}Harvard-Smithsonian Center for Astrophysics, Cambridge, MA, USA\\
\llap{$^h$}Max-Planck-Institut f\"ur Radioastronomie (MPIfR), Auf dem H\"ugel 69, D-53121 Bonn, Germany\\
E-mail: \email{algaba@astro.snu.ac.kr}}

\abstract{The flat spectrum radio quasar 4C 38.41 showed a significant increase of its radio flux density during the period 2012 March - 2015 August which correlates with gamma-ray flaring activity. Multi-frequency simultaneous VLBI observations were conducted as part of the interferometric monitoring of gamma-ray bright active galactic nuclei (iMOGABA) program and supplemented with additional monitoring observations at various bands across the electromagnetic spectrum. The epochs of the maxima for the two largest gamma-ray flares coincide with the ejection of two respective new VLBI components and the evolution of the physical properties seem to be in agreement with the shock-in-jet model. Derived synchrotron self absorption magnetic fields, of the order of 0.1 mG, do not seem to dramatically change during the flares, and are much smaller, by a factor 10,000, than the estimated equipartition magnetic fields, indicating that the source of the flare may be associated with a particle dominated emitting region.}

\FullConference{14th European VLBI Network Symposium \& Users Meeting (EVN 2018)\\
		8-11 October 2018\\
		Granada, Spain}

\begin{document}

\section{Introduction}

The source 4C~38.41 is a flat-spectrum radio quasar (FSRQ) at a redshift z=1.813 \cite{Hewett10}. Strong variability in its radio flux density has been observed \cite{Kuhr81,Spangler81,Aller92}, and superluminal motion with jet velocities of up to $\sim30$~c has been detected \cite{Lister13}. The $\gamma$-ray flares observed by the \emph{Fermi}/LAT (Large Area Telescope) in 2009-2010 were associated with an emerging component from the core downstream of the jet at the 43~GHz \cite{Jorstad11}. On the other hand, a large outburst observed in 2011 was explained geometrically to be due to variations of the Doppler factor owing to changes in the viewing angle \cite{Raiteri12}.

The quasar 4C~38.41 showed a significant increase of its radio flux density during the period 2012 March - 2015 August. This
provides an excellent opportunity to perform a multi--band analysis to study the location of the $\gamma$-ray flares, the emission mechanisms responsible for their origin, and their connection with the evolution of radio flares in great detail.

\section{Observations}

Multi-frequency simultaneous VLBI observations at 22, 43, 86 and 129~GHz were conducted as part of the interferometric monitoring of gamma-ray bright active galactic nuclei (iMOGABA) program \cite{Lee16}. We obtained additional data for 4C~38.41 from the 43~GHz VLBA Boston University (BU) Blazar program to obtain a better cadence and coverage of the light curve at at 43~GHz. Radio observations were supplemented with archival data from the 15~GHz radio monitoring program with the 40~m telescope at the Owens Valley Radio Observatory (OVRO) and the 225~GHz Submillimeter Array (SMA) monitoring program. 

In addition, we obtained additional monitoring observations at various bands across the electromagnetic spectrum. Optical data was obtained from the Steward Observatory of the University of Arizona, which makes reduced V-band photometric data publicly available. X--ray data in the energy range 0.2-10 keV from the \emph{Swift}-XRT monitoring of \emph{Fermi}-LAT sources of interest are publicly available thanks to support from the Fermi GI program and the \emph{Swift} Team. To investigate the $\gamma$-ray flux variations at GeV energies, we used the \emph{Fermi}-LAT data observed in survey mode.

\section{Results and Discussion}

The light curve of 4C~38.41 indicates high flux densities, more than twice as large as usual, observed in radio bands between MJD 56200 and MJD 56700 (2012 March - 2015 August). The associated optical, X-ray, and $\gamma$-ray fluxes seem to follow a similar trend, although for optical and X-rays, the poor sampling complicates the comparison. Moreover, a clear flux peak seen in these bands at MJD 57050 is not visible in any radio band. See \cite{Algaba18a} for details.

\begin{figure}
\vspace{-0.5cm}
\includegraphics[scale=0.63]{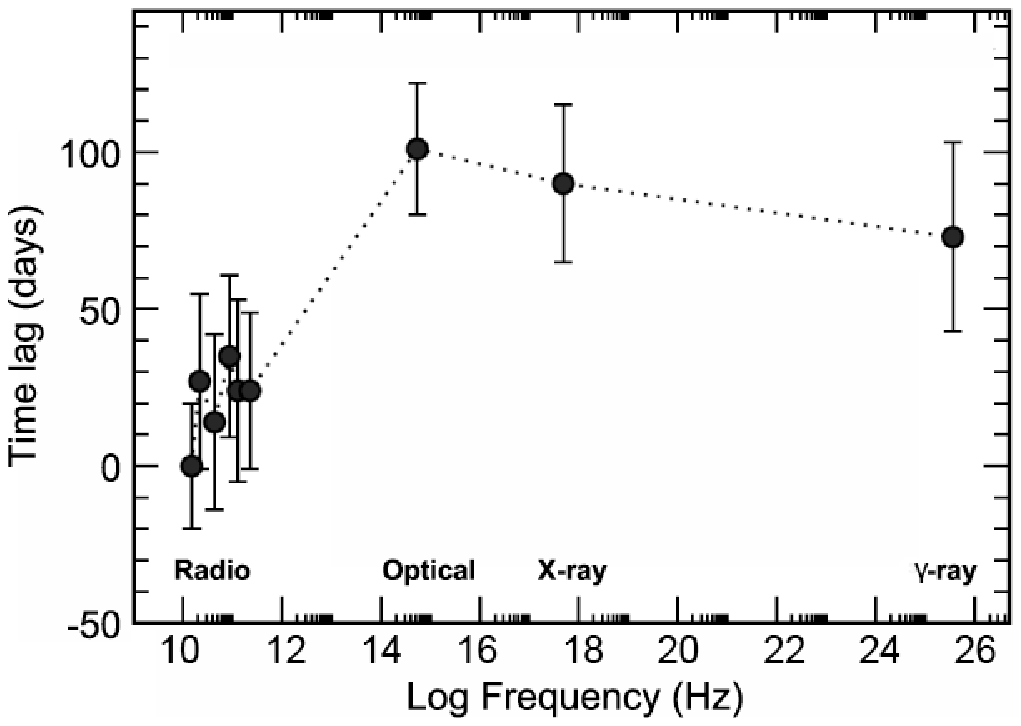}
\includegraphics[scale=0.63]{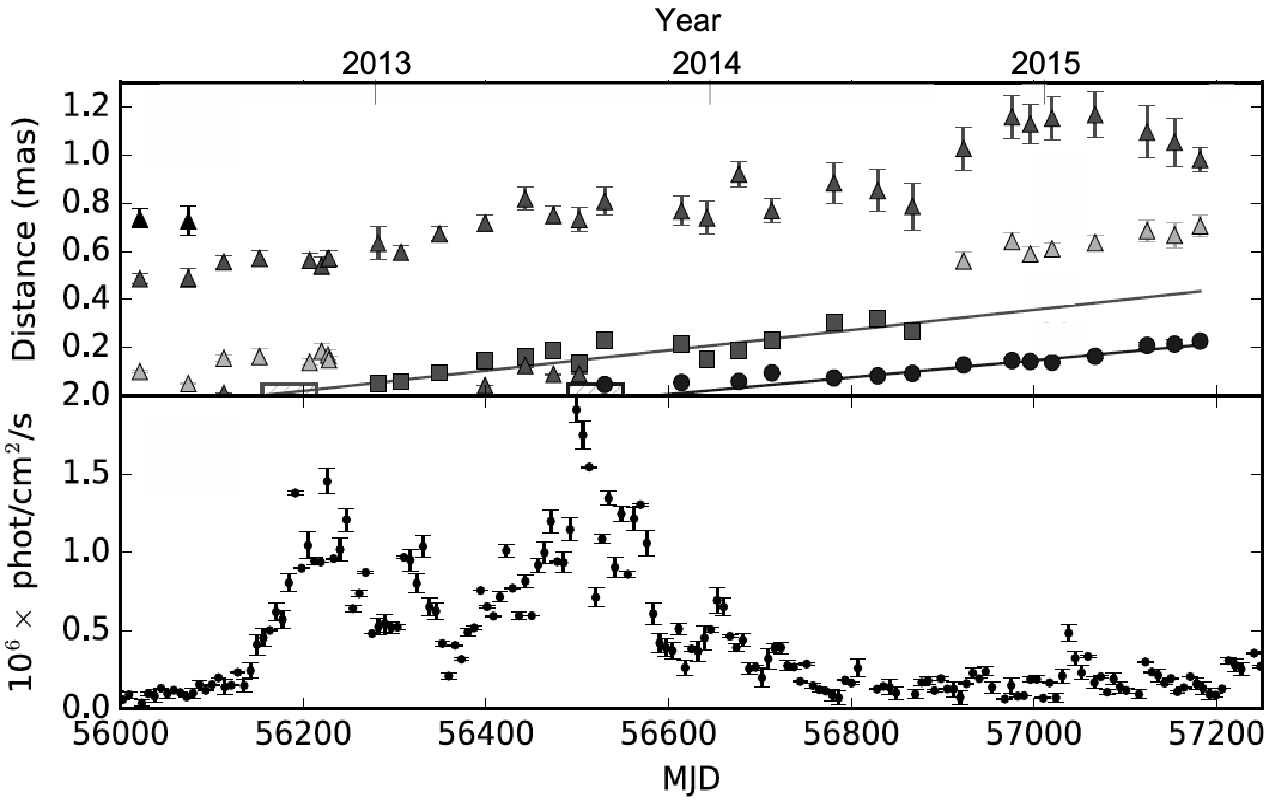}
\vspace{-0.6cm}
\caption{\textbf{Left:} Time lag between Radio GHz and other bands obtained via the DCF. \textbf{Right:} Top panel shows the evolution of distance of the components from the VLBI core as resolved with BU 43~GHz data. Straight lines are a linear fit for the components speed, with the boxes indicating the uncertainty in the fitted ejection date. The bottom panel shows the flux density of $\gamma$-rays for comparison.}
\label{fig1}
\end{figure}

Cross-correlation analysis shows the flux in the different bands to be significantly correlated, with the possible exception of optical bands, where the correlation, while still present, is not statistically significant (<95\%). Analysis of the discrete correlation function (DCF) suggests time lags smaller than the uncertainty in the peak of the DCF among radio frequencies, as well as among high energies (optical, X-rays, and $\gamma$-rays), whereas a time lag of about 70-90 days is found between radio and high-energy bands (see Figure \ref{fig1}, left), suggesting that the emissions at high energies and in radio bands are produced in two different jet regions, with the $\gamma$-rays located at $1\pm13$~pc and radio emission at $40\pm13$~pc from the jet apex. This is in agreement with the shock-in-jet model \cite{MarscherGear85} where a new component is ejected from the core and travels thought the jet.

\begin{figure}
\includegraphics[scale=0.71,trim={0cm 0.5cm 0cm 1.5cm},clip]{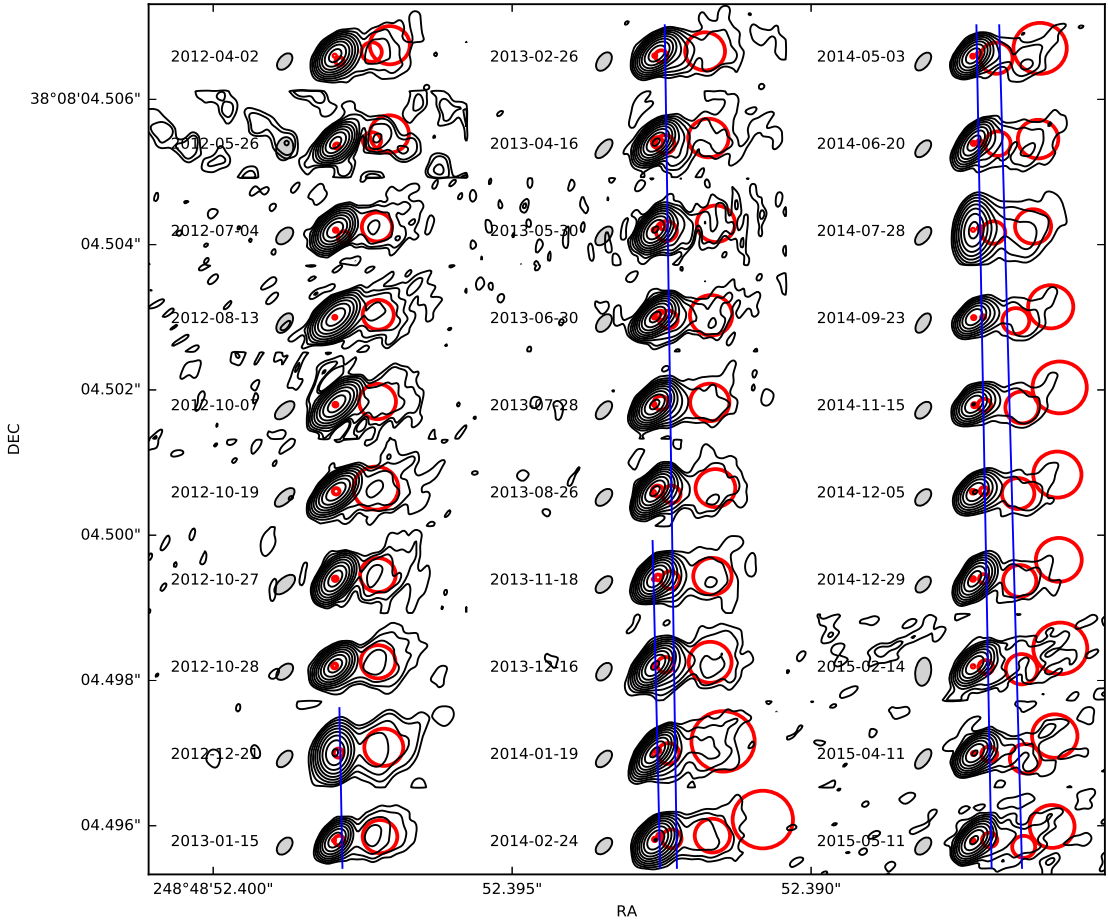}
\vspace{-1.0cm}
\caption{BU 43 GHz VLBI maps. Contours start at 3 times the median rms, equal to 2.5~mJy. The observation date and the beam size are shown on the left of each map. Red circles indicate the circular model-fitted components. Blue lines indicate components displacement; note that actual time difference between epochs is not equal and the lines serve for eye guidance only.}
\label{fig2}
\end{figure}

Supporting this model, resolved components by the BU 43 GHz VLBI data are found to be moving away from the core. Two of them, with constant speeds of $10.2\pm0.8$ and $11.7\pm1.6$~c, have extrapolated ejection epochs MJD=$56520\pm30$ and MJD=$56185\pm30$, respectively, which fall well within the epochs for which the largest $\gamma$-rays were observed (Figure \ref{fig1}, right). This seems to indicate that the $\gamma$-ray flaring is tightly associated with the ejection of these components. On the other hand, there are no radio structural changes associated with the dimmer $\gamma$-ray flares (see \cite{Algaba18b} for details; detailed multi-epoch BU VLBA data at 43~GHz are shown in Figure\ref{fig2}). The reported flaring activity in the source can simply be explained by radiative processes having a constant Doppler factor.

\begin{figure}
\includegraphics[scale=0.53]{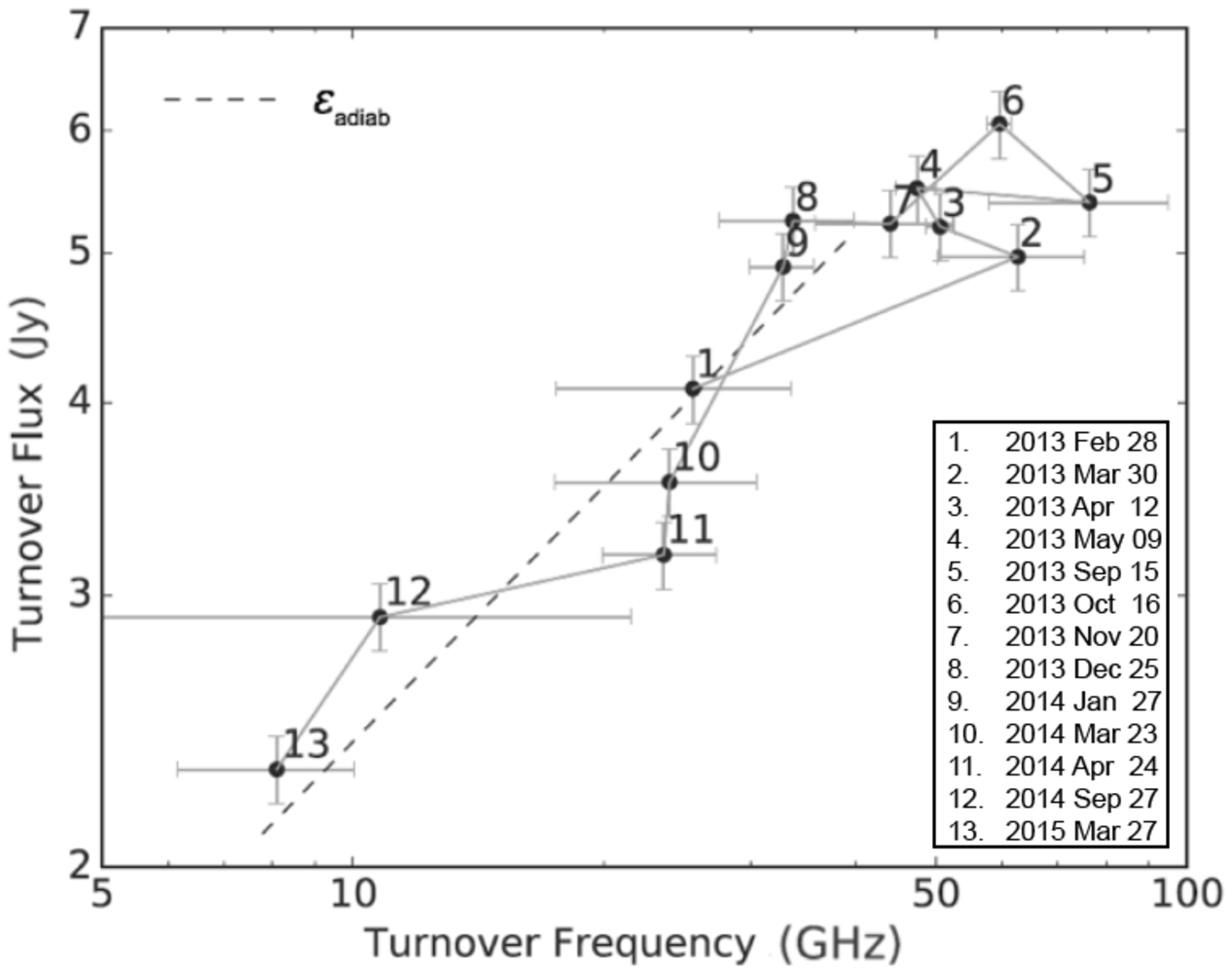}
\includegraphics[scale=0.45]{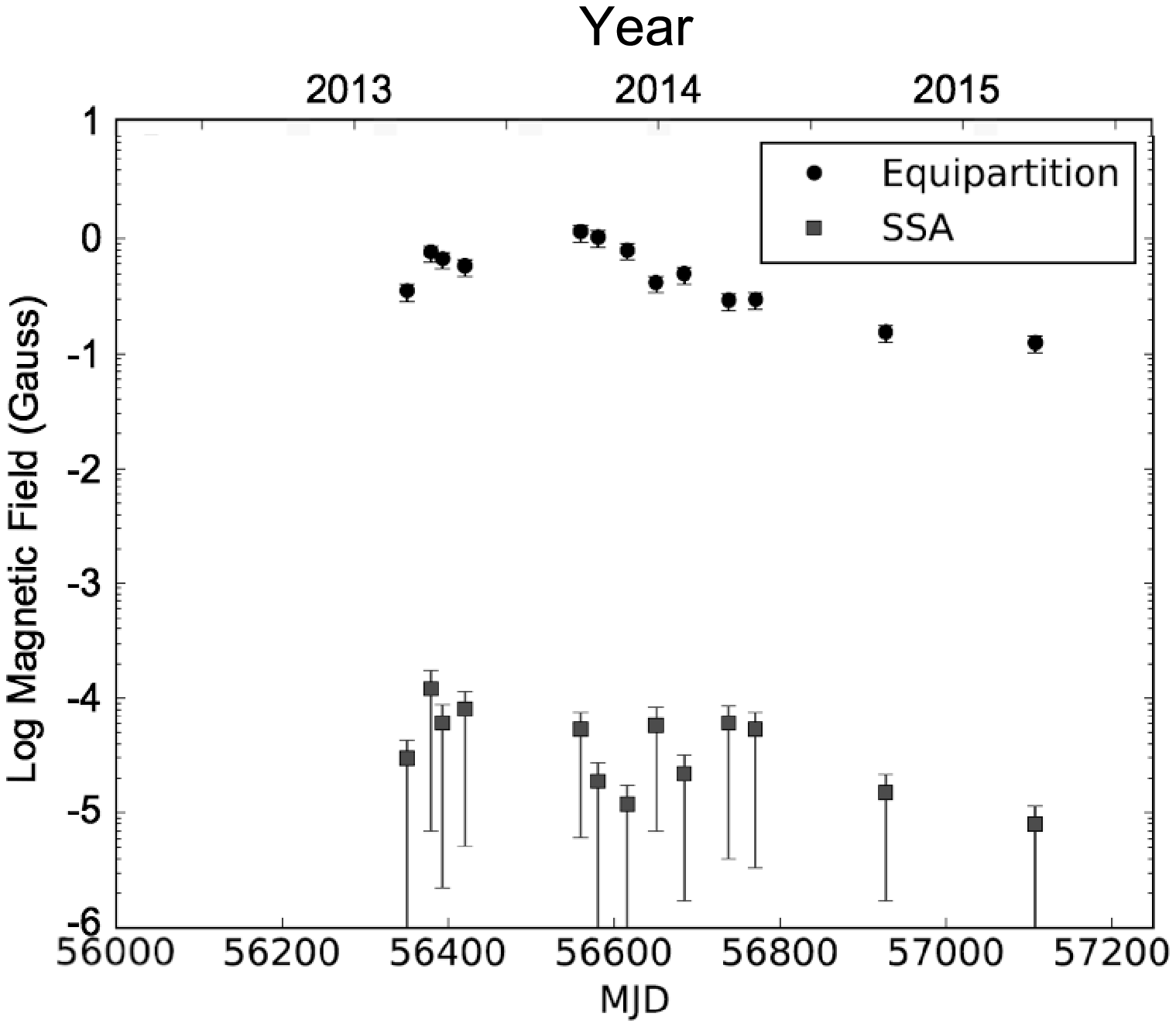}
\caption{\textbf{Left:} Turnover flux - turnover frequency diagram. Each number refers to the various epochs for which turnover values were obtained, as shown in the bottom left corner. Note that the time gap between epochs may be different. Dashed line correspond to the fit for points 8-13. \textbf{Right:} Equipartition (black circles) and synchrotron self-absorption (grey squares) magnetic fields. }
\label{fig3}
\end{figure}

We investigated the radio spectral energy distribution (SED) as follows: first we concentrated on the iMOGABA simultaneous data. Then we searched for the closest epochs for OVRO at 15~GHz and SMA at 225~GHz. We considered only these epochs within two weeks (14 days) of the iMOGABA data for the SED. To calculate the turnover frequency, we restricted our study to epochs where at least five data points could be obtained. Our analysis indicates that the turnover frequency shifts from few GHz to few tens of GHz after the more luminous, long-lived $\gamma$-ray flares occur \cite{Algaba18b}. The evolution of the flare in the turnover frequency-turnover flux density plane (Figure \ref{fig3}, left) shows an initial complicated pattern for the Compton and synchrotron losses stages due to the overlap of the effects from two interleaved flares, while the adiabatic loss stage is very clear, with a slope $\epsilon_{\mbox{adiab}}=0.6\pm0.1$, in agreement with the shock-in-jet model in \cite{MarscherGear85}.

Following the fitted parameters for the BU 43~GHz VLBI core, we estimated the magnetic field strength via both synchrotron self-absorption and equipartition considerations (Figure \ref{fig3}, right). The magnetic field estimated via synchrotron self-absorption does not significantly vary over time and is of the order of 0.1~mG, smaller by a factor $10^4$ than the magnetic field strength estimated using equipartition arguments. These two findings suggest that the emitting region of the flares is particle dominated.

\section{Conclusions}

4C~38.41 showed an increase of its radio flux density correlated with $\gamma$-ray flares with radio enhancement following that of high energies by about 70-90 days. This phenomena can be associated with the ejecta of new components from a particle dominated region, becoming visible as radiation reaches optically thin regions. Follow-up of the components location, speed, flux density and turnover frequency shows that emission is in agreement with the shock-in-jet model adiabatically expanding with a constant Doppler factor.

\end{document}